\documentstyle[prl,aps,twocolumn]{revtex}

\begin{document}
\twocolumn[\hsize\textwidth\columnwidth\hsize\csname
@twocolumnfalse\endcsname

\title{Open Charm Enhancement in Pb+Pb Collisions at SPS}

\author{
M.I. Gorenstein$^{a,b}$,
%\footnote{E--mail: goren@th.physik.uni-frankfurt.de},
A.P. Kostyuk$^{a,b}$,
%\footnote{E--mail: kostyuk@th.physik.uni-frankfurt.de},
H. St\"ocker$^{a}$
%\footnote{E--mail: stoecker@th.physik.uni-frankfurt.de}
and 
W. Greiner$^{a}$
%\footnote{E--mail: greiner@th.physik.uni-frankfurt.de}
}

\address{
$^a$ Institut f\"ur Theoretische Physik, Universit\"at  Frankfurt,
Germany}

\address{$^b$ Bogolyubov Institute for Theoretical Physics,
Kiev, Ukraine}

\date{\today}
\maketitle

\begin{abstract}
The statistical coalescence model for the production of open and hidden
charm
is considered within the canonical ensemble formulation.
The data for the $J/\psi$ multiplicity in  Pb+Pb collisions at
158~A$\cdot$GeV are used for the model  prediction of the
open charm yield. We find a strong enhancement of the open charm
production, 
by a factor of about 2--4, over the standard hard-collision
model extrapolation from
nucleon-nucleon to
nucleus-nucleus collisions.   
A possible mechanism of the open charm enhancement in A+A collisions at
the SPS energies is proposed. 
\end{abstract}


]

The charmonium states $J/\psi$ and $\psi^{\prime}$ have been measured in
nucleus-nucleus (A+A)
collisions at CERN SPS over the last 15 years by the NA38 and NA50
Collaborations.
This experimental program was motivated by a suggestion
 \cite{Satz1} to use the $J/\psi$ 
as a probe of the state of matter created in the early stage
of the collision. In this approach a significant suppression
of $J/\psi$ production relative to Drell--Yan lepton pairs
is predicted when going 
from peripheral to central Pb+Pb interactions at 158~A$\cdot$GeV.
This is originally
attributed to the formation of a quark-gluon plasma,
but could be also explained in microscopic hadron models as secondary
collision effects (see \cite{Sp} and references therein).

The statistical approach, formulated  in
Ref.\cite{Ga1}, assumes that
$J/\psi$ mesons are created at hadronization according to the available
hadronic phase-space similar to other (lighter) hadrons.
We call this a hadron gas (HG) model. 
The HG model offers a natural explanation of the
proportionality of the $J/\psi$ and pion yields in nuclear collisions.

Recently a picture of the $J/\psi$ creation via 
coalescence (recombination)
of charmed quarks was developed within several model approaches
\cite{Kabana:2000sd,Br1,Le:00,Ra:00}. 
Similar to the HG model \cite{Ga1}, 
the charmonium states are assumed 
to be formed at the hadronization stage. 
However, they are produced as a coalescence of created earlier
$c$ and $\overline{c}$ quarks and therefore 
the multiplicities of open and hidden charm
hadrons are connected in the coalescence models.
In Ref.~\cite{Br1} the
charm quark-antiquark pairs are assumed to be created at the early
stage of A+A collision and the average number of
directly produced $c$-$\overline{c}$ pairs,
$N^{dir}_{c\overline{c}}$, 
is fixed by the model consideration based on the hard scattering 
approach.
This requires the introduction of a new parameter in the HG approach --
the charm enhancement factor $\gamma_c$ \cite{Br1}. 
This is analogous to the
introduction of the strangeness suppression factor $\gamma_s$ \cite{Raf1}
in the HG model, where the total strangeness observed is smaller than its
thermal equilibrium
value. Within this approach the open charm hadron yield is enhanced
by a factor $\gamma_c$ and 
the charmonium yield by a factor $\gamma_c^2$ in comparison
with the equilibrium HG predictions.

As the total number of charmed hadrons is expected to be smaller than
unity, even for the most central Pb+Pb collisions at the
SPS energy, an exact
charm conservation within the canonical ensemble (c.e.) should be
imposed. This has been done in Ref.\cite{Go:00}.
Note also that the c.e. formulation was successfully used in
Ref.\cite{BecD} to
calculate the open charm hadron abundances in $e^+e^-$ collisions
%at $\sqrt{s}=91.2$~GeV 
with an experimental input  of the total open charm
production. 

In this letter 
we consider the
c.e. formulation \cite{Go:00}  of the statistical coalescence model
\cite{Br1}.
The experimental value
of the $J/\psi$ multiplicity $\langle J/\psi\rangle$ will be used
to predict the open charm yield within the statistical coalescence model.
We find a
strong enhancement of the open charm production,
by a factor of about 3, over the standard hard-collision   
model extrapolation from
nucleon-nucleon to
A+A collisions.

\vspace{0.3cm}
In the grand canonical ensemble (g.c.e.) the thermal multiplicities of
both open charm and
charmonium states are given as (Bose and Fermi effects are negligible):
\begin{equation}\label{gce}
N_j~=~ \frac{d_j
V~e^{\mu_j/T}}{2\pi^2}~T~m^2_j~K_2\left(\frac{m_j}{T}\right)~,
%\cong~d_j~V~e^{\mu_j/T}~\left(\frac{m_jT}{2\pi}\right)^{3/2}~\exp\left(-
%\frac{m_j}{T}\right)~,
\end{equation}
where $V$ and $T$ correspond to the volume\footnote{To avoid further
complications we use ideal HG formulae and neglect excluded volume
corrections.} and
temperature
of HG system, $m_j$, $d_j$ denote particle masses and degeneracy
factors and $K_2$ is the modified Bessel function. The particle chemical
potential $\mu_j$ in Eq.(\ref{gce})
is defined as
\begin{equation}\label{mui}
\mu_j~=~b_j\mu_B~+~s_j\mu_S~+~c_j\mu_C~,
\end{equation}
where $b_j,s_j,c_j$ denote the baryonic number strangeness and
charm of particle $j$. The baryonic chemical potential $\mu_B$
regulates the baryonic density of the HG system whereas
strange $\mu_S$ and charm $\mu_C$ chemical potentials should be found
from the requirement of zero value for the total strangeness and charm
in the system (in our consideration we neglect small effects
of a non-zero electrical chemical potential). 

In the c.e. formulation (i.e. when the requirement of zero "charm charge" 
of the HG is used in the exact form) the thermal charmonium
multiplicities are still given by Eq.(\ref{gce}) as charmonium states
have zero 
charm charge.
The multiplicities (\ref{gce}) of {\it open} charm hadrons 
will, however, be multiplied by the additional 
'canonical suppression' factor (see e.g. \cite{Go1}).
This suppression factor is the same for all individual open 
single charm states.
It leads to the total open charm multiplicity $N_O^{ce}$ in the c.e.: 
\begin{equation}\label{ce}
 N_O^{ce}~=~ N_O~\frac{I_1(N_O)}{I_0(N_O)}~,
\end{equation}
where
$N_O$ is the total g.c.e.
multiplicity of all open charm and anticharm mesons and (anti)baryons 
calculated with Eq.(\ref{gce})
and $I_0,I_1$ are the modified Bessel functions.   
For large open charm multiplicity $N_O>>1$ one finds
$I_1(N_O)/I_0(N_O)\rightarrow 1$ and therefore $N_O^{ce}\rightarrow
N_O$,
i.e. the g.c.e. and c.e. results coincide. For $N_O<<1$ one has
$I_1(N_O)/I_0(N_O)\cong N_O/2$ and
$N^{ce}_O\cong N_O \cdot N_O/2$,
therefore, $N_O^{ce}$ is strongly suppressed (like $N_O/2$, $N_O << 1$)
in comparison to the g.c.e. result, $N_O$ . 

Assuming the
presence of the charm enhancement factor $\gamma_c$ 
the statistical coalescence model within the c.e. 
is formulated as:
\begin{equation}\label{Ncc1}
N^{dir}_{c\overline{c}} = \frac{1}{2}~\gamma_c~N_O~\frac{I_1(\gamma_c
N_O)}{I_0(\gamma_cN_O)}
+ \gamma_c^2~N_{H}~,
\end{equation}
when $N_{H}$
is the total   
(thermal) multiplicity of particles with hidden charm.
Therefore, the baryonic number, strangeness and electric
charge of the HG system are treated in our approach according to the
g.c.e. but charm is considered in the c.e.
formulation where the exact charge conservation is imposed.

We will proceed with Eq.(\ref{Ncc1}) in the
following way.
As the  $\langle J/\psi \rangle$ multiplicities
can be extracted from the NA50 data
on Pb+Pb collisions at 158 A$\cdot$GeV for
different values of $N_p$,
we start from the requirement:
\begin{equation}\label{Npsi}
\langle J/\psi \rangle~= ~\gamma_c^2~N_{J/\psi}^{tot}~,
\end{equation}
to fix the $\gamma_c$ factor.
In Eq.(\ref{Npsi}) the total $J/\psi$ thermal multiplicity
is calculated as 
\begin{eqnarray}\label{dec}
N_{J/\psi}^{tot} &=& N_{J/\psi} + R(\psi^{\prime})N_{\psi^{\prime}}\\
& & + R(\chi_1)N_{\chi_1} + R(\chi_2)N_{\chi_2}~,
\end{eqnarray}
where $N_{J/\psi}$, $N_{\psi^{\prime}}$, $N_{\chi_1}$, $N_{\chi_2}$
are given by Eq.(\ref{gce}) and $R(\psi^{\prime})\cong 0.54$,
$R(\chi_1)\cong 0.27$, $R(\chi_2)\cong 0.14$ are the decay branching
ratios of the excited charmonium states into $J/\psi$.  
Eq.(\ref{Ncc1}) will be used then to calculate the value of
$N^{dir}_{c\overline{c}}$. This value will be considered as a prediction
of the statistical coalescence model:  the open charm yield has not yet
been measured
in  Pb+Pb collisions at SPS.

A reliable extraction
of the
$J/\psi$ yields from the
published data appears to be
non-trivial\footnote{We are thankful 
to  M.~Ga\'zdzicki and  K. Redlich  for
useful comments.}.
% Analysis done in  Ref.\cite{GaJ} suggests
% an approximately linear increase of $\langle J/\psi \rangle$
% with $N_p$. 
The results for $\langle J/\psi \rangle$ presented
in Ref.\cite{Ga1} were evaluated
from the data of the NA50 Collaboration \cite{NA50new} using the
procedure described in \cite{APP}.
These results are presented in Table 1 to be used in the present
analysis.

We use the
set of the chemical freeze-out parameters \cite{Br2}:
\begin{equation}\label{set}
T~=~168~{\rm MeV},~~ \mu_B~=~266~{\rm MeV}~.
\end{equation}
They are fixed
by fitting the  HG model 
to the hadron yield data
in Pb+Pb collisions at 158 A$\cdot$GeV
 (the inclusion of open charm and charmonium
states does not modify the rest of the hadron yields).
For a fixed number of participants, $N_p$, 
the volume $V$ is calculated from $N_p=Vn_B$,
where $n_B=n_B(T,\mu_B)$ is the baryonic density calculated in
the g.c.e.. 
With the chemical freeze-out parameters (\ref{set}) we 
find the $N_{J/\psi}^{tot}$ and $N_O$ values using Eq.(\ref{gce}).
Then we calculate $\gamma_c$
from Eq.(\ref{Npsi}). Finally,
$N^{dir}_{c\overline{c}}$ is calculated
from Eq.(\ref{Ncc1}). 
Note that $T\cong170$~MeV leads to the HG value of the thermal ratio
of $\langle \psi^{\prime} \rangle /\langle J/\psi \rangle \cong 0.04$
in agreement with data \cite{ratio} at $N_p >100$.
This fact was first noticed in Ref.\cite{Sh}. 

\vspace{0.3cm}
The model results for central 
Pb+Pb interactions at 158~A$\cdot$GeV ($N_p=100\div 400$)
are presented in Table 1.
Assuming 
that $N^{dir}_{c\overline{c}}$ scales as $N_p^{\alpha}$
we find $\alpha \cong 1.7$. 
This value is larger than $\alpha \cong 4/3$ expected
in the hard-collision model.

pQCD inspired models suggest values of
$N_{c\bar{c}}^{dir} = 0.15\div 0.3$ in central Pb+Pb collisions at 158
A$\cdot$GeV (a value of $N_{c\bar{c}}^{dir}\cong 0.17$ is
estimated in Ref.\cite{Br1}). The results presented in Table 1
correspond to an enhancement of the open charm production
by a factor of about 2--4. Although the values of $N_{J/\psi}^{tot}$,
$N_O$ and $\gamma_c$ are rather sensitive to the temperature parameter, the
model predictions for $N_{c\bar{c}}^{dir}$ remain essentially unchanged 
when 
$T=170 \pm 10$ MeV is used.
Note that a 
recent analysis of the dimuon spectrum measured in central Pb+Pb
collisions at 158 A$\cdot$GeV by NA50 Collaboration \cite{NA50}
suggests a significant enhancement of dilepton
production in the intermediate mass region (1.5$\div$2.5 GeV) over
the standard sources.
The primary\footnote{Alternative explanations are also suggested 
\cite{Spieles:1998wz,Rapp:2000zw,Gallmeister:2000dj}. } 
interpretation attributes this observation to the
enhanced production of open charm \cite{NA50}:
about 3 times above the pQCD prediction
for the open charm yield in Pb+Pb collisions at SPS.
This value is in agreement with our model analysis.
Moreover, the analysis of the centrality dependence of the dimuon
excess was done in Ref.\cite{Kabana:2000sd} assuming that additional 
dimuon pairs come entirely from decays of $D$ and $\bar{D}$. The open 
charm yield was found to be proportional to $N_p^{1.7}$. This precisely 
coincides with our result.   

A possible source of the open charm enhancement in A+A
collisions
with respect to the direct extrapolation from p+p data may
be the
broadening of the phase space available for the open charm due to the
presence
of the hadron and/or the quark-gluon medium:

Let a $c\bar{c}$ pair  with an invariant mass below the open
charm threshold, $2M_D\cong 3.7$~GeV, be created in a p+p collision.
In the vacuum it cannot hadronize to an open charm hadron pair
and thus it has to be transformed into non-charmed states to respect
energy-momentum conservation. In A+A collisions,  however,
such a pair 
has a chance to
shift its invariant mass above the open charm threshold
due to secondary interactions with hadron
and/or quark-gluon
medium, thus
resulting in additional open charm production.
The strongest enhancement of the open charm production should be expected 
in the quark-gluon plasma. Due to the Debye screening, $c$ and $\bar{c}$
behave like free particles. Therefore, almost all subthreshold 
$c\bar{c}$-pairs created in the hard collisions hadronize into open 
charm mesons and baryons at the later stage of the A+A reaction.
This effect can be substantial at moderate center-of-mass energies
per
nucleon, while at higher energies the threshold effects
becomes less important. A quantitative estimate of this effect is 
under way \cite{prep}. 
The above mechanism does not alter
the production of direct
Drell-Yan pairs:
both the threshold effects and the interaction with the nuclear
medium are negligible in that case.

\vspace{0.3cm}
In conclusion, the statistical coalescence model with exact charm 
conservation has been formulated. The canonical ensemble suppression
effects
are important for the thermal open charm yield even at $N_p\cong 400$. 
These effects become
crucial when the number of participants $N_p$ decreases.
From the $J/\psi$ multiplicity data in  Pb+Pb collisions at
158~A$\cdot$GeV the
open charm yield is predicted: $$N_{c\bar{c}}^{dir} = 0.6\div 0.7$$ in
 central collisions. This is about a factor 3  
above the pQCD prediction
for the open charm yield in Pb+Pb collisions at SPS.
The statistical coalescence model
predicts also an $N_p$ dependence of $N_{c\bar{c}}^{dir}\sim N_p^{1.7}$,
which is stronger than the standard dependence, $N_p^{4/3}$, in
the hard-collision model.
These predictions of the statistical coalescence model
can be tested
in the near future at the CERN SPS, where measurements of the open charm are
planned. 

\acknowledgments
%{\bf  Acknowledgments.}  

The authors
are thankful to F. Becattini, P. Braun-Munzinger, K.A.~Bugaev, M. Ga\'zdzicki,
L. Gerland, L.~McLerran,
I.N.~Mishustin, G.C.~Nayak,  K.~Redlich and J.~Stachel for
numerous valuable comments and
discussions.
We acknowledge the financial support of GSI, DFG, BMBF and DAAD,
 Germany.
The research described in this publication was made possible in part by
Award \# UP1-2119 of the U.S. Civilian Research and Development
Foundation for the Independent States of the Former Soviet Union
(CRDF).

\newpage
\begin{table}[p]
\begin{center}
Table 1 \\
\vspace{0.3cm}
\begin{tabular}{|c|c|c|c|c|c|}
%\hline 
\rule[-3mm]{0mm}{10mm}
  &$\langle J/\psi \rangle \cdot 10^4$ & $N_{J/\psi}^{tot}\cdot
10^4$
      &$N_{O}$  & $\gamma_c$ &
           { $N_{c\bar{c}}^{dir}$ }  \\                                            
$N_{p}$ & NA50 data \cite{NA50new} &  &  &  &  \\
 & Compil. \cite{Ga1} & Eq.(\ref{dec})  &   
    &Eq.(\ref{Npsi})& Eq.(\ref{Ncc1}) \\
\hline
 100 & 2.2 $\pm$ 0.2 & 0.56 & 0.26  & 2.0   & 0.066    \\
 200 & 3.9 $\pm$ 0.2 & 1.1\ & 0.52 & 1.9   & 0.21\    \\
 300 & 6.4 $\pm$ 0.6 & 1.7\ & 0.79 & 2.0   & 0.46\    \\
 360 & 6.9 $\pm$ 0.7 & 2.0\ & 0.94  & 1.9   &  0.57\    \\
%\hline
\end{tabular}
\end{center} 
\end{table}
%\vspace{0.3cm}

\end{document}